\documentclass[10pt,a4paper,twoside]{article}
\usepackage{epsfig}
\usepackage{baltlat6}
\usepackage{array}
\usepackage{here}
\begin{document}
\ \
\vspace{0.5mm}
\setcounter{page}{1}

\titlehead{Baltic Astronomy, vol.\,xx, xxx--xxx, 2016}

\titleb{ON UTMOST MULTIPLICITY OF HIERARCHICAL STELLAR SYSTEMS}

\begin{authorl}
\authorb{Y. M. Gebrehiwot}{1},
\authorb{D. A. Kovaleva}{2},
\authorb{A. Yu. Kniazev}{3,4,5,6},
\authorb{O. Yu. Malkov}{2},
\authorb{N.~A.~Skvortsov}{7},
\authorb{A. V. Karchevsky}{8},
\authorb{S. B. Tessema}{1},
\authorb{A. O. Zhukov}{2,5,9}
\end{authorl}

\begin{addressl}
\addressb{1}{Entoto Observatory and Research Center,
Astronomy and Astrophysics Research Division,
P.O. Box 33679 Addis
Ababa, Ethiopia; yikdema16@gmail.com}
\addressb{2}{Institute of Astronomy, Russian Acad. Sci.,
48 Pyatnitskaya St., Moscow 119017, Russia}
\addressb{3}{South African Astronomical Observatory, PO Box 9, 7935 Observatory,
South Africa}
\addressb{4}{Southern African Large Telescope Foundation, PO Box 9, 7935 Observatory,
Cape Town, South Africa}
\addressb{5}{Sternberg Astronomical Institute, Moscow State University,
13 Universitetskij Prosp., Moscow 119992, Russia}
\addressb{6}{Special Astrophysical Observatory, Russian Acad. Sci., Nizhnij Arkhyz,
Karachai-Cherkessian Republic 369167, Russia}
\addressb{7}{Institute of Informatics Problems, Russian Acad. Sci.,
44-2 Vavilova St., Moscow 119333, Russia}
\addressb{8}{Faculty of Physics, M.\,V. Lomonosov Moscow State University,
Moscow 119992, Russia}
\addressb{9}{Federal Center of Expertise and Analysis,
Russian Ministry of Education and Science, 33-4 Talalikhina St.,
Moscow 109316, Russia}

\end{addressl}

\submitb{Received: 2016 xxx xx; accepted: 20xx xxx xx}

\begin{summary}
According to theoretical considerations, multiplicity of
hierarchical stellar systems can reach, depending on masses and
orbital parameters, several hundred, while observational data
confirm existence of at most septuple (seven-component) systems.
In this study, we cross-match very high multiplicity (six and more
components) stellar systems in modern catalogues of visual double
and multiple stars, to find candidates to hierarchical systems
among them. After cross-matching with catalogues of closer
binaries (eclipsing, spectroscopic, etc.), some of their
components were found to be binary/multiple themselves, which
increases the system's degree of multiplicity. Optical pairs,
known from literature or filtered by the authors, are flagged and
excluded from the statistics. We have compiled a list of
potentially very high multiplicity hierarchical systems that
contains 10~objects. Their multiplicity does not exceed 12, and we
discuss a number of ways to explain the lack of extremely high
multiplicity systems.
\end{summary}


\resthead{On utmost multiplicity of hierarchical stellar systems}
{Gebrehiwot et al.}

\sectionb{1}{INTRODUCTION}

Data on stellar multiplicity is important as a constraint on the
problem of the formation and evolution of the Galactic stellar
population. On the other hand, statistics of stellar multiplicity,
i.e. the number of components, is poorly known, especially for
multiplicities higher than three or four, and many questions still
remain unresolved (see, e.g., the recent review by Duch\^ene \&
Kraus 2013).

The maximum number of components in a hierarchical multiple system
depends on the number of hierarchy levels and can be estimated
from theoretical considerations. A system is dynamically stable
if, in a case of circular orbits, the outer orbital period exceeds
the inner orbital period by a factor of five. For eccentric
orbits, this factor is larger, increasing as $\propto (1-e)^3$
(Tokovinin 2004). The mean outer/inner ratio for the semi-major
axis and period is 20 and 70, respectively. On the other hand, the
number of levels in hierarchical stellar systems is limited by the
tidal action of regular gravitational field of the Galaxy,
gravitational perturbations from passing stars, and stochastic
encounters with giant molecular clouds (see, e.g., Jiang \&
Tremaine 2010). Surdin (2001) demonstrated that, in these
circumstances, the number of levels can reach values of 8 or 9,
depending on masses and orbital parameters of the components. In
the case of maximum dense ``packing'' of components in the system,
hierarchical systems with 256 to 512 components can be produced.

On the other hand, there is no evidence to prove the existence of
any hierarchical system having multiplicity of seven or higher.
The most comprehensive catalogue of multiple systems (Tokovinin
1997) contains about 1350 hierarchical systems of multiplicity
three to seven, and among the two catalogued septuple systems
(AR~Cas and $\nu$~Sco), at least the former one is a young
cluster, i.e. is not necessarily hierarchical. This statistics is
in a sharp contrast with the theoretical estimates given above. To
eliminate this inconsistency, it is necessary to use additional
sources of information, namely modern catalogues of double stars.

\sectionb{2}{CATALOGUES OF DOUBLE AND MULTIPLE SYSTEMS}

\begin{table}[!t]
\begin{center}
\vbox{\footnotesize\tabcolsep=3pt
\parbox[c]{124mm}{\baselineskip=10pt
{\smallbf\ \ Table~1.}{\small\ Principal catalogues of visual double and
multiple systems\lstrut}}
\begin{tabular}{|l|r|l|}
\hline
Catalogue, abbreviation, reference & C, P, S & M     \hstrut\lstrut\\
\hline
The Washington Double Star         & 249280  & 2-32 \\
Catalog (WDS, Mason et al. 2016)   & 133966  &      \\
                                   & 115314  &      \\
\hline
Catalogue of Components of         & 105837  & 1-18 \\
Double and Multiple Stars          &  56513  &      \\
(CCDM, Dommanget \& Nys 2002)      &  49325  &      \\
\hline
Tycho Double Star Catalogue        & 103259  & 1-11 \\
(TDSC, Fabricius et al. 2002)      &  37978  &      \\
                                   &  64869  &      \\
\hline
\end{tabular}
}
\end{center}
\vskip-2mm
\begin{small}
C, P, S are numbers of catalogued components, pairs, and systems,
respectively,

M is multiplicity of catalogued systems.
\end{small}
\end{table}

Principal modern catalogues of visual double stars contain systems
of much higher multiplicity than seven (see Table~1). Actually,
WDS contains several systems of even higher multiplicity than
indicated in Table~1. They represent either results of searches
for sub-stellar companions to nearby stars by high-contrast and
high-angular-resolution imaging, where at least some of the
objects are background stars (WDS~17505$-$0603, 65 objects;
WDS~19062$-$0453 = $\lambda$ Aql, 107 objects) or results of
speckle interferometric observations of stars in nebulae
(WDS~05387$-$6906 = 30~Dor = Tarantula Nebula, 68 objects;
WDS~05353$-$0523 = $\theta^1$~Ori = Trapezium cluster in Orion
nebula, 39 objects) or miniclusters / common proper motion groups
(WDS~19147+1918, WDS~20315+3347, WDS~13447$-$6348,
WDS~18354$-$3122, WDS~23061+6356, 38 to 44 objects per system).

\begin{figure}
\includegraphics[width=7.0cm]{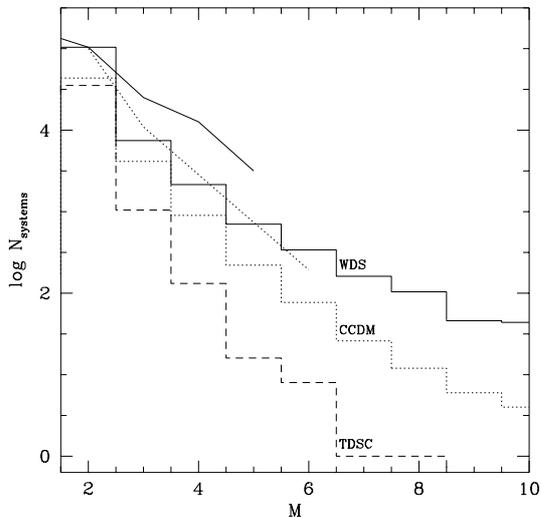}
\caption{Observational distribution of stellar systems on
multiplicity M. Histograms represent distributions from principal
catalogues of double and multiple stars: WDS (solid), CCDM
(dotted), TDSC (dashed). Curves represent published statistics on
hierarchical stellar systems; dotted curve: Orlov \& Titov (1994)
and Tokovinin (2001); solid curve: Tokovinin (2014). The curves
are normalized on the number of binaries in WDS.}
\label{fig:multiplicity}
\end{figure}

Note also in brackets that CCDM and TDSC contain some systems of
multiplicity one. In the former case, this concerns astrometric
binaries (with an invisible secondary component detected by its
gravitational influence), while TDSC contains a fair amount of
stars that the Tycho mission failed to resolve into components.

It is instructive to plot the distribution of catalogued stellar
systems on their multiplicity and compare it to the observational
data. Tokovinin (2001) presented statistics of catalogued multiple
systems in the form $N_i/N_{i-1}=$ 0.11, 0.22, 0.20, 0.36 for
$i=3$, 4, 5, 6, respectively, where $N_i$ is the number of systems
of the $i$th multiplicity. These results are in a good accordance
with conclusions made by Orlov \& Titov (1994) in their study of
multiple objects in the immediate ($d \le 25$~pc) solar vicinity.
Later Tokovinin (2014) studied hierarchical systems among F and G
dwarfs in the Solar neighborhood and found the fraction of
multiple systems with 1, 2, 3, ... components to be 54:33:8:4:1.
These results are plotted in Fig.~1. It can be seen that the most
complete catalogue, WDS, satisfactory images the Orlov \& Titov
(1994) and Tokovinin (2001) distributions, while the newer and
deeper study by Tokovinin (2014) demonstrates a surplus of triple
and higher-multiplity systems in comparison with the catalogued
systems (or, conversely, WDS contains superfluous, obviously
optical, double stars).

The listed catalogues of visual binaries contain various data for
evidently overlapping sets of objects, and no one of them contains
all known visual systems. Thus, to use the complete dataset, it
was necessary to cross-match these catalogues, i.e. to gather all
available information on visual binary stars in a single list. A
comprehensive set of visual binaries using data from the current
versions of the three listed catalogues was compiled by Isaeva et
al. (2015). However, as further analysis has shown, the applied
cross-matching procedure worked quite well for systems of
multiplicity about five or six, but often failed to correctly
cross-identify components in systems of higher multiplicity, due
to high spatial density of objects.

\sectionb{3}{VERY HIGH MULTIPLICITY SYSTEMS: PROCEDURE, RESULTS,\\ AND DISCUSSION}

To compile a list of candidates to hierarchical stellar systems of
maximum multiplicity (and estimate the value of this maximum
multiplicity), as well as to finally solve the problem of
cross-identification of multiple systems, we have performed a
semi-manual identification of systems of multiplicity six and more
in principal catalogues of visual double and multiple systems (see
Table~1). The total number of such systems is 551. 175 of them are
included in WDS only.
The remaining 395 systems are included in more than one catalogue
and, consequently, their components need cross-matching (the
systems themselves were cross-matched by Isaeva et al. 2015 and
analyzed in Kovaleva et al. 2015a).

Compiling the list of very high multiplicity systems, we were
flagging optical pairs. The information about non-physical nature
of a pair can be found in WDS and the textual Notes to WDS. We
have also applied the criterion to select optical pairs suggested
by Poveda et al. (1982), which revealed additional optical
objects. The result can be seen in Fig.~2 (gray bars).

\begin{figure}
\includegraphics[width=8.0cm]{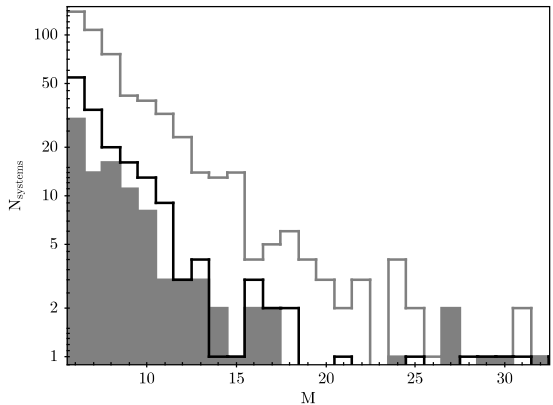}
\caption{Observational distribution of stellar systems on
multiplicity M. The gray contour represents all systems. Gray bars
represent all systems except optical pairs. The black contour
represents all systems except optical pairs, but with the addition
of closer, photometrically unresolved components (see text for
details). Systems of multiplicity higher than 32 are discussed in
the beginning of Section~2 and are not shown here.}
\label{fig:optical}
\end{figure}

Photometrically unresolved binarity of some components can
increase actual multiplicity of a system. In order to take this
into account, we have cross-matched our systems with lists of
closer binaries (orbital, interferometric, spectroscopic,
eclipsing, X-ray systems, radio pulsars) using the Binary star
database, BDB (Kaygorodov et al. 2012, Kovaleva et al. 2015b).
Besides, indication of hidden binarity can sometimes be found in
WDS Notes (34 cases). Information on sub-components of our very
high multiplicity systems was found in catalogues of orbital (77
pairs), interferometric (425 pairs), spectroscopic (52), and
eclipsing (16) binaries. The resulting distribution can be seen in
Fig.~2 (the black contour).

Finally, we have excluded from our statistics those pairs that
have no clear indication on their physical binarity, according to
WDS and the textual Notes to WDS. As a result, we have compiled a
list of 10 so-called ``confirmed'' systems of multiplicity six and
higher. The list contains all systems included in the WDS, CCDM,
and TDSC catalogues, and thus it is the most comprehensive list of
stellar systems of multiplicity six and more. We provide extensive
cross-identifications for components, pairs, and systems included
in the list. We add data on photometrically unresolved binaries,
taken from catalogues of closer pairs (spectroscopic, eclipsing,
etc.) and flag optical pairs.

\begin{table}[!t]
\begin{center}
\vbox{\footnotesize\tabcolsep=3pt
\parbox[c]{124mm}{\baselineskip=10pt
{\smallbf\ \ Table~2.}{\small\ Multiple system statistics\lstrut}}
\begin{tabular}{rrrr|rrrr|rrrr}
\hline
   M & N1   & N2 & N3& M & N1  & N2 &N3& M & N1  & N2& N3 \\
\hline
   6 & 138  & 54 & 5 &15 & 14  & 1  &- &24 &  4  & - &-\\
   7 & 107  & 34 & 3 &16 &  4  & 3  &- &25 &  2  & 1 &-\\
   8 & 76   & 20 & - &17 &  5  & 2  &- &26 &  1  & - &-\\
   9 & 42   & 16 & - &18 &  6  & 2  &- &27 &  1  & - &-\\
  10 & 39   & 13 & 1 &19 &  4  & -  &- &28 &  -  & 1 &-\\
  11 & 32   &  9 & - &20 &  3  & -  &- &29 &  -  & 1 &-\\
  12 & 23   &  3 & 1 &21 &  2  & 1  &- &30 &  1  & 1 &-\\
  13 & 14   &  4 & - &22 &  3  & -  &- &31 &  2  & 1 &-\\
  14 & 13   &  1 & - &23 &  -  & -  &- &32 &  1  & - &-\\
\hline
\end{tabular}
}
\end{center}
\vskip-2mm
\begin{small}
M: multiplicity of systems; N1: number of candidate systems; N2:
number of prospective systems; N3: number of confirmed systems.
\end{small}
\end{table}

The final statistics is presented in Table~2. Column N1 contains
the number of candidates to systems of multiplicity M (gray
contour in Fig.~2). Column N2 contains the number of candidates to
systems of multiplicity M, without optical pairs (black contour in
Fig.~2). Column N3 contains the number of confirmed systems of
multiplicity M.

\begin{table}[!t]
\begin{center}
\vbox{\footnotesize\tabcolsep=3pt
\parbox[c]{124mm}{\baselineskip=10pt
{\smallbf\ \ Table~3.}{\small\ Systems of highest prospective multiplicity\lstrut}}
\begin{tabular}{lrrr}
\hline
            ID   & M1   & Opt &  M2 \\
\hline
WDS 23061+6356   &   31 &  8 &  1 \\
WDS 17378$-$1315 &   30 &  2 &  1 \\
WDS 10174$-$5354 &   29 &  3 &  2 \\
WDS 10451$-$5941 &   28 &  6 &  1 \\
WDS 15326$-$5221 &   25 &  0 &  1 \\
WDS 17457$-$2900 &   21 &  0 &  7 \\
WDS 01030+6914   &   18 &  1 &  1 \\
WDS 05353$-$0522 &   18 &  0 &  1 \\
\hline
\end{tabular}
}
\end{center}
\vskip-2mm
\begin{small}
M1: number of components without optical ones; Opt: number of
optical pairs; M2: possibly confident hierarchical multiplicity.
\end{small}
\end{table}

The highest-multiplicity systems are listed in Table~3. For each
system, the number of components (M1), number of optical
components (Opt), and number of confidently hierarchical
components (M2) are given. The system WDS~17457$-$2900
demonstrates the highest value of possible hierarchical
multiplicity (7), while possible multiplicity of several other
systems (WDS~23061+6356, WDS~17378$-$1315, WDS~10174$-$5354)
reaches higher values, but it should be confirmed by observations.

It can be seen that these values are still far from those expected
from theoretical predictions. Several possible ways can be
considered to explain such a mismatch.

First, the theoretical possibility to construct a system with 8--9
hierarchy levels is based on purely geometrical considerations and
does not necessarily mean that physical conditions in a
protostellar cloud can permit to construct such a system.
Consecutive fragmentation of a large contracting interstellar
cloud is needed for a very high multiplicity hierarchical system
to be born.

Also, very wide binaries ($a \ge 100$~AU) are so weakly bound that
they can be effectively disturbed, even disrupted, by extremely
weak perturbations from inhomogeneities in the Galactic potential
due to stars, molecular clouds, dark objects, or large-scale
tides. Thus, the outermost components of a very high multiplicity
hierarchical system will probably not survive on their orbits and
leave the system.

Finally, Fig.~1 demonstrates that we probably underestimate hidden
multiplicity of stellar systems, and the number of photometrically
unresolved components is much higher than catalogued data predict.

\sectionb{4}{CONCLUSIONS}

To explain inconsistency in stellar multiplicity between rather
high values predicted by theoretical considerations and
observational lack of systems with multiplicity higher than six,
we have studied principal catalogues of visual double stars: WDS,
CCDM and TDSC. They contain data on very high multiplicity (up to
30 components and more), thought not necessarily hierarchical,
systems, including moving groups and (mini-)clusters. To collect
all available information on these systems, it was first necessary
to make a thorough and accurate cross-matching of their components
in the catalogues. Optical pairs, when known or assumed from the
probability filter, were flagged and eliminated from the
statistics, and information on photometrically unresolved
sub-components was added.

Principal results of the current study are the following.
\begin{itemize}
\item a cross-identification catalogue of 551 stellar systems of
multiplicity six and more; \item a list of systems, candidates to
utmost multiple hierarchical systems; \item a procedure for
cross-matching components of very high multiplicity systems (i.e.,
in crowded stellar fields), which also can be used for
identification of objects in future surveys of binary/multiple
stars (Gaia, LSST).
\end{itemize}

\thanks{
The work was partly supported by the Presidium of the Russian
Academy of Sciences program ``Leading Scientific Schools Support''
9951.2016.2 and by the Russian Foundation for Basic Research
grants 15-02-04053 and 16-07-01162.
A.\,K. acknowledges support from the National Research Foundation
of South Africa and by the Russian Science Foundation (project no.
14-50-00043). This research has made use of the VizieR catalogue
access tool (the original description of the VizieR service was
published in A\&AS 143, 23), the SIMBAD database, operated at the
Centre de Donn\'ees astronomiques de Strasbourg, the Washington
Double Star Catalog maintained at the U.S. Naval Observatory, and
NASA's Astrophysics Data System Bibliographic Services. }

\References

\refb Dommanget J., Nys O. 2002, VizieR On-line Data Catalog: I/274

\refb Duch\^ene G., Kraus A. 2013, Annu. Rev. Astron. Astrophys.
51, 269

\refb Fabricius C., H{\o}g E., Makarov V. V. et al. 2002, A\&A, 384, 180

\refb Isaeva A. A., Kovaleva D. A., Malkov O. Yu. 2015, Baltic
Astronomy 24, 157

\refb Jiang Y.-F., Tremaine S. 2010, MNRAS 401, 977

\refb Kaygorodov P., Debray B., Kolesnikov N. et al. 2012, Baltic Astronomy, 21, 309

\refb Kovaleva D. A., Malkov O. Yu., Yungelson L. R. et al. 2015a,
Baltic Astronomy 24, 367

\refb Kovaleva D. A., Kaygorodov P. V., Malkov O. Yu. et al.
2015b, Astronomy \& Computing 11, 119

\refb Mason B. D., Wycoff G. L., Hartkopf W. I. et al. 2016, VizieR On-line Data Catalog: B/wds

\refb Orlov V. V., Titov O. A. 1994, Astronomy Reports 38, 462

\refb Poveda A., Allen C., Parrao L. 1982, ApJ 258, 589

\refb Surdin V. G. 2001, ASP Conf. Ser. 228, 568

\refb Tokovinin A. A. 1997, A\&AS 124, 75

\refb Tokovinin A. A. 2001, in {\em Proc. IAU Symp. 200, The
Formation of Binary Stars}, eds. H. Zinnecker, R. D. Mathieu,
Potsdam, April 2000, ASP, San Francisco, USA, 84

\refb Tokovinin A. A. 2004 in Rev. Mex. Astron. Astrof. Conf.
Ser., Ed. by C. Allen and C. Scarfe (Instituto de Astronom\'\i a,
UNAM, Mexico) 21, 7

\refb Tokovinin A. A. 2014, AJ 147, 87

\end{document}